\RequirePackage{fix-cm}
\documentclass[twocolumn,epjc3]{svjour3}  
\usepackage{rotating}
\usepackage{xcolor}
\usepackage{amsmath}
\usepackage{hyphenat}
\smartqed  

\usepackage{tikz}
\usetikzlibrary{arrows.meta}
\usetikzlibrary{positioning,calc}
\RequirePackage{graphicx}
 \usepackage{booktabs}
\journalname{Eur. Phys. J. C}

\newcommand{\pc}{pseu\hyp{}do\hyp{}complex}

\begin{document} 

\title{Curvature Regularization and Dynamical Vacuum Structure in
  Pseudo-Complex General Relativity
}


\author{Fridolin Weber\thanksref{e1,addr1,addr2}
        \and
        Peter O. Hess\thanksref{e2,addr3,addr4} \and 
        Cesar A. Zen Vasconcellos\thanksref{e3,addr5,addr6}
}

\thankstext{e1}{e-mail: fweber@sdsu.edu}
\thankstext{e2}{e-mail: hess@nucleares.unam.mx}
\thankstext{e3}{e-mail: cesaraugustozenvasconcellos@gmail.com}

\institute{Department of Physics, San Diego State University, San
  Diego, CA, 92182, USA \label{addr1}
  \and
Department of Physics, University of California San Diego, 
La Jolla, CA, 92093, USA \label{addr2}
\and
 Instituto de Ciencias Nucleares, Universidad Nacional Autónoma de México, Mexico City, 04510, Mexico \label{addr3}
\and
Frankfurt Institute for Advanced Studies (FIAS), 
J. W. von Goethe Universität, Frankfurt am Main, Germany \label{addr4}
\and
International Center for Relativistic Astrophysics Network (ICRANet), 
Pescara, Italy \label{addr5}
\and
Universidade Federal do Rio Grande do Sul, Porto Alegre, RS, 91501-970, Brazil \label{addr6}       
}

\date{Received: date / Accepted: date}

\maketitle

\begin{abstract}
Whether classical spacetime remains well-defined at arbitrarily high
curvature is a central question. In general relativity, singularities
signal the breakdown of the classical description, motivating
modifications of spacetime's short-distance structure. Pseudo-complex
general relativity (pcGR) extends spacetime geometry to pseudo-complex
coordinates, naturally introducing two metric sectors, the physical
metric $g_{\mu\nu}$ and an auxiliary field $f_{\mu\nu}$. The magnitude
of the pseudo-imaginary component defines the invariant acceleration
scale $a_0$, a direct consequence of the pseudo-complex structure. The
simultaneity condition, requiring both idempotent sectors to satisfy
the pseudo-complex Einstein equations independently, fixes the
auxiliary field algebraically, with no new propagating degrees of
freedom, and sets a lower bound on the lapse that regularizes
curvature.
 The regularization is achieved through the combined effect
of the lapse-gap condition \( e^{\nu(r)} > a_0 \), which ensures the
radial metric component remains nondegenerate, together with the
regularity conditions at the areal-radius origin, \( B(0)=1 \) and \(
B'(0)=0 \), which emerge naturally from the pseudo-complex
geometry. These conditions jointly eliminate the Schwarzschild-type
curvature divergence, rather than the gap condition acting alone.

In cosmology, we show that the pseudo-complex field equations admit a
formal reduction to an effective evolution equation for the vacuum
component $\rho_{\rm pc}(t)$: the auxiliary functions $f_0$ and $f_2$,
together with $\ddot a$, are formally eliminated within the restricted
parametrization adopted in the reduction, yielding a second-order
evolution equation in which no auxiliary functions remain explicitly.

The reduced evolution equation admits asymptotic solutions whose
behavior is determined by the pseudo-complex geometry. In particular,
it admits late-time solutions approaching an effective constant vacuum
energy, suggesting a geometric origin for dark energy with scale set
by $a_0$. pcGR thus realizes, through purely geometric
conditions, bounded curvature, intrinsic cutoff scales, dynamical
vacuum, and absence of global symmetries, while features depending on
quantum spectra have no classical analog.
\end{abstract}

\section{Introduction}\label{sec:intro}

General relativity (GR) provides a highly successful geometric description of
gravitation across a wide range of scales. Nevertheless, classical solutions of
the Einstein equations generically develop curvature singularities in regimes of
strong gravitational collapse, where curvature invariants diverge and geodesic
evolution breaks down. Such singularities signal a limitation of the classical
theory and motivate the search for frameworks that remain well defined at high
curvature \cite{Bardeen1968,AyonBeatoGarcia1998,Modesto2006,Nicolini2006}.
For a recent review of regular black-hole constructions and their underlying
mechanisms, see \cite{Lan2023RegularBHReview}.

A variety of approaches address this problem by introducing new degrees of
freedom, nonlocal dynamics, or quantum corrections. An alternative possibility
is that the admissible class of spacetime geometries is itself restricted by
additional geometric consistency conditions, without enlarging the dynamical
field content. In such a framework, singular behavior may be avoided not through
new interactions, but through constraints on the allowed configurations of the
metric.

Pseudo-complex general relativity, first developed by Hess and
Greiner \cite{HessGreiner2009} (see also
\cite{HessSchaferGreiner2015}), provides a concrete realization of
this idea. In pcGR, spacetime coordinates are promoted to
pseudo-complex variables $X^\mu = x^\mu + I a^\mu$, introducing an
invariant acceleration scale and an auxiliary geometric sector. The
pseudo-complex algebra, with its idempotent decomposition $e_\pm =
\frac{1}{2}(1 \pm I)$, allows any pseudo-complex quantity to be split
into two independent sectors. This structure enables the introduction
of a second metric component $f_{\mu\nu}$ without adding new dynamical
degrees of freedom: the auxiliary sector is fixed algebraically by the
simultaneity condition, requiring both idempotent sectors to satisfy
the Einstein equations independently. Unlike typical modified gravity
theories, which introduce new dynamical fields with their own kinetic
terms and wave equations, pcGR introduces no new propagating modes.
The theory modifies the space of admissible spacetimes through
constraints, not through new dynamics.
In symmetric configurations, this modification alters the radial and lapse
functions so that,
  together with the regular-center conditions discussed
below, curvature invariants remain finite at $r=0$, yielding nonsingular
solutions without invoking new matter fields or quantum corrections.
The resulting regularization, together with the emergence of an
intrinsic cutoff scale, is therefore purely geometric in origin.

The extension introduces an invariant acceleration scale \( a_0 \), a
direct consequence of the pseudo-complex structure, not a free
parameter,  whose precise definition and relation to the physical
scale \( a_{0,\rm phys} \) are given in Sec.~\ref{sec:2}.

A central question underlying this work is whether the restrictions
imposed by the pseudo-complex extension are merely sufficient
conditions for regularity in specific symmetric configurations, or
whether they reflect a more general principle: that the space of
admissible spacetime geometries is itself constrained by algebraic
consistency conditions. We therefore analyze the geometric
consequences of the pseudo-complex constraint structure not as a
phenomenological regularization scheme, but as a probe into the
structural conditions under which classical geometry can remain
well-defined without additional dynamical input. Our focus is on
isolating which features, i.e., bounded curvature, intrinsic cutoff scales,
dynamical vacuum behavior, arise directly from the constraint
structure alone, and which require supplementary assumptions or
extensions beyond the pseudo-complex framework.

The structure of this paper is as follows. In Sec.~\ref{sec:2} we
review the pseudo-complex geometric framework and field equations. In
Sec.~\ref{sec:pcgr_solutions} we analyze spherically symmetric
solutions and derive the conditions under which curvature
regularization occurs.  In Sec.~\ref{sec:cosmo} we examine
cosmological solutions and
derive the effective evolution equation for
the vacuum component.
In Sec.~\ref{sec:swampland} we present a
structural comparison with Swampland-inspired constraints, and in
Sec.~\ref{sec:spec.outlook} we discuss possible connections to broader
geometric frameworks. We conclude in Sec.~\ref{sec:conclusion}.

\section{Pseudo-Complex Geometry and Field Equations}\label{sec:2}

The pseudo-complex extension of GR is constructed on the algebra of
pseudo-complex numbers, where the coordinates assume the form
\(
X^\mu=x^\mu_{(1)}+Ix^\mu_{(2)}
\),
with \(I^2=1\). This algebraic structure was originally introduced to
address the singularities of classical GR while preserving the geometric
nature of the theory. Rather than modifying the Einstein equations or
introducing additional matter fields, the pseudo-complex approach
extends the algebraic structure of spacetime itself. Its central feature
is the introduction of an auxiliary metric sector without adding new
propagating degrees of freedom.

The pseudo-complex shift
\(
X^\mu=x^\mu+Ia^\mu
\),
where \(a^\mu\) is a constant pseudo-complex vector, naturally
introduces an invariant acceleration scale. Its magnitude defines the
physical inverse-length scale \(a_{0,\rm phys}\), which is a geometric
property of the pseudo-complex structure rather than an externally
introduced parameter. In the limit
\(a_{0,\rm phys}\rightarrow0\),
the two idempotent sectors coincide and standard general relativity is
recovered. Throughout this work we distinguish between the physical
scale \(a_{0,\rm phys}\) and the dimensionless parameter \(a_0\)
appearing in the metric, which are related by
\[
a_0\equiv \ell\,a_{0,\rm phys},
\]
where \(\ell\) is the length scale used to nondimensionalize the radial
coordinate. The role of \(a_0\) in regularizing the geometry is discussed
in Sec.~\ref{sec:pcgr_solutions}.
  
The pseudo-complex approach includes both field-theoretic and
gravitational formulations. An early pseudo-complex field theory,
building on the work of Schuller, was developed in
\cite{HessGreiner2007}. The extension to pseudo-complex general
relativity was introduced in \cite{HessGreiner2009}, where the
Schwarzschild solution was obtained, followed by Kerr and
Reissner--Nordstr\"om solutions in \cite{Caspar2012}.
Subsequent applications include neutron-star models
\cite{Rodriguez2014}, cosmological pcFLRW solutions in which the
dark-energy component emerges geometrically from the pseudo-complex
structure \cite{HessMaghlaouiGreiner2010,Maghlaoui2026},
ray-tracing studies consistent with EHT observations
\cite{Schonenbach2013,Schonenbach2014,Schonenbach2016},
gravitational-wave constraints from the GWTC-3 catalog
\cite{NielsenBirnholtz2017,Maimon2025}, and connections to
Ho\v{r}ava--Lifshitz gravity
\cite{HessZenVasconcellosHadjimichef2025}. Collectively, these results
show that pcGR provides a phenomenologically consistent framework that
remains compatible with current observations while making testable
predictions for future experiments.

The simultaneity condition, that the \pc{} Einstein equations must
hold independently in both idempotent sectors \(e_+\) and \(e_-\),
follows directly from the algebraic structure of the pseudo-complex
numbers and is discussed in Refs.~\cite{HessGreiner2009,HessSchaferGreiner2015}.

Pseudo-complex numbers form a commutative ring defined by \(I^2=+1\). They
admit the idempotent decomposition
\cite{HessGreiner2009,HessSchaferGreiner2015}
\begin{equation}
e_\pm=\frac{1}{2}(1\pm I), \qquad
e_\pm^2=e_\pm, \qquad
e_+e_-=0,
\end{equation}
which allows any pseudo-complex quantity to be written as
\begin{equation}
Z=Z_+e_+ + Z_-e_-.
\end{equation}
This decomposition defines two mutually orthogonal sectors that may be
treated independently, subject to the consistency conditions imposed by
the pseudo-complex structure.

The differential structure
\begin{equation}
dX^\mu = dx^\mu + I\,da^\mu
\end{equation}
induces the pseudo-complex line element
\begin{equation}
d\omega^2 = G_{\mu\nu}\,dX^\mu dX^\nu,
\end{equation}
where the pseudo-complex metric is
\begin{equation}
G_{\mu\nu}=g_{\mu\nu}+If_{\mu\nu}.
\label{eq:pc-metric}
\end{equation}
Here \(g_{\mu\nu}\) denotes the physical spacetime metric, while
\(f_{\mu\nu}\) is an auxiliary symmetric tensor encoding the
pseudo-complex extension.

Projecting onto the idempotent basis gives
\begin{equation}
d\omega^2=d\omega_+^2\,e_+ + d\omega_-^2\,e_-,
\end{equation}
with
\begin{equation}
d\omega_\pm^2=
(g_{\mu\nu}\pm f_{\mu\nu})
(dx^\mu\pm da^\mu)
(dx^\nu\pm da^\nu).
\end{equation}
The two sectors therefore define a pair of effective geometries related
by the pseudo-complex constraint structure.

Lorentzian signature in the physical sector is preserved provided that
\(f_{\mu\nu}\) remains sufficiently small compared with \(g_{\mu\nu}\),
and the GR limit is recovered as \(f_{\mu\nu}\rightarrow0\). Throughout
this work we adopt the metric signature \((+,-,-,-)\).

The invariant acceleration scale introduced above enters the metric
through the pseudo-complex extension and governs the separation between
the two idempotent sectors. In the dimensionless metric expressions used
below, we define
\begin{equation}
a_0 \equiv \ell\, a_{0,\rm phys},
\end{equation}
where $\ell$ is the length scale used to nondimensionalize the radial
coordinate. Throughout this paper, $a_0$ denotes the corresponding
dimensionless parameter appearing in the metric, while
$a_{0,\rm phys}$ denotes the associated physical inverse-length scale.
As shown in Sec.~\ref{sec:pcgr_solutions}, this scale determines the
lapse-gap condition responsible for curvature regularization.

\subsection{Field Equations}\label{ssec:field_equations}

Replacing $g_{\mu\nu}$ by the pseudo-complex metric
(\ref{eq:pc-metric}) in the Einstein--Hilbert functional yields the
pseudo-complex Einstein equations,
\begin{equation}
R_{\mu\nu}(\mathcal{G}) - \frac{1}{2} R(\mathcal{G})\, \mathcal{G}_{\mu\nu}
= \kappa\, T_{\mu\nu}(\mathcal{G}) \, ,
\label{eq:pceinstein}
\end{equation}
where $\kappa = 8\pi G$, $R_{\mu\nu}(\mathcal{G})$ and
$R(\mathcal{G})$ denote the Ricci tensor and scalar curvature
constructed from $\mathcal{G}_{\mu\nu}$.

The pseudo-complex structure requires that Eq.~(\ref{eq:pceinstein}) be
satisfied simultaneously in both idempotent sectors. Projecting onto the
idempotent basis therefore yields two coupled Einstein-type systems,
\begin{eqnarray}
R_{\mu\nu}^{(+)} - \frac{1}{2} R^{(+)} g_{\mu\nu}^{(+)} &=& \kappa\,
T_{\mu\nu}^{(+)},  \\
R_{\mu\nu}^{(-)} - \frac{1}{2} R^{(-)} g_{\mu\nu}^{(-)} &=& \kappa\,
T_{\mu\nu}^{(-)}\, ,
\end{eqnarray}
with
\begin{equation}
g_{\mu\nu}^{(\pm)} = g_{\mu\nu} \pm f_{\mu\nu}, \qquad
T_{\mu\nu}^{(\pm)} = T_{\mu\nu}(g_{\mu\nu}^{(\pm)})\, .
\end{equation}
Because both sectors must satisfy these equations simultaneously,
their mutual consistency determines $f_{\mu\nu}$ algebraically, up to
integration constants fixed by symmetry.  The simultaneity condition,
that the pseudo-complex Einstein equations must hold independently in
both idempotent sectors \(e_+\) and \(e_-\), follows from the
algebraic structure of the pseudo-complex numbers and is derived in
the foundational work \cite{HessGreiner2009}. This simultaneity
condition is not merely a formal requirement; it enters explicitly in
the derivation of the curvature bounds in \ref{sec:appendixA}, where
the independence of the two idempotent sectors ensures that the
denominator \( A(r)^2 - a_0^2 \) controls the entire invariant
spectrum (see Eqs.~\eqref{eq:A.5}--\eqref{eq:A.7} and the surrounding
discussion).

This structure implies that the auxiliary tensor $f_{\mu\nu}$ does not obey an
independent evolution equation. Instead, it is fixed by constraint relations
linking the two effective metrics $g_{\mu\nu}^{(\pm)}$. The resulting system is
therefore analogous to constrained formulations in which auxiliary fields are
determined algebraically once the dynamical variables are specified.

As a consequence, the number of propagating degrees of freedom coincides with
that of general relativity at the level of the classical field equations.
By construction, the pseudo-complex extension modifies the admissible
class of spacetime solutions through algebraic constraints. The
auxiliary field $f_{\mu\nu}$ does not represent an independent
dynamical degree of freedom but is determined by the pseudo-complex
constraint structure.

Since the field equations are modified, Birkhoff's theorem does not hold
in its standard form. In the present work, we restrict attention to
static, spherically symmetric configurations. The existence of static
solutions therefore follows from this symmetry assumption together with
the pseudo-complex field equations, rather than from a general
uniqueness theorem.

In the weak-field regime, pseudo-complex general relativity reduces to general
relativity whenever the dimensionless correction
$a_0^2/e^{2\nu(r)}$ is small, ensuring compatibility with
post-Newtonian bounds and solar-system observations in the appropriate
parameter regime \cite{HessSchaferGreiner2015}.

\subsection{Regularization of Curvature}

A central consequence of the pseudo-complex constraint structure is the
regularization of curvature in high-sym\-metry settings. In general relativity,
solutions such as the Schwarz\-schild metric develop divergent curvature
invariants at $r=0$. A standard diagnostic is the Kretsch\-mann scalar,
\begin{equation}
\lim_{r \to 0} R_{\mu\nu\rho\sigma} R^{\mu\nu\rho\sigma} = \infty .
\end{equation}

In pseudo-complex general relativity, the invariant acceleration scale $a_0$
restricts the behavior of the lapse and radial functions in such a way that this
divergence is avoided. In particular, the pseudo-complex structure enforces a
lower bound on the lapse function, $e^{\nu(r)} \ge a_0$, which prevents the
vanishing behavior responsible for the Schwarzschild singularity. As a result,
curvature invariants remain finite at $r=0$ in the spherically symmetric
solutions constructed in Ref.~\cite{HessGreiner2009}.

This regularization mechanism is geometric rather than dynamical. No additional
matter fields, nonlocal terms, or quantum corrections are required. Instead, the
pseudo-complex extension restricts the allowed spacetime configurations through
algebraic consistency conditions, and these constraints are sufficient to
control the curvature, provided the additional consistency
conditions at the areal-radius origin are satisfied, as discussed in detail in
Sec.~\ref{sec:pcgr_solutions}.


\section{Spherically Symmetric Solutions in pcGR}\label{sec:pcgr_solutions}

Spherically symmetric configurations provide a setting in which the
geometric consequences of the pseudo-complex constraint structure can be
analyzed explicitly. In particular, they allow a direct demonstration of
how bounded curvature and an intrinsic cutoff scale arise from the
modified relation between metric components.

In general relativity, Birkhoff's theorem implies that any spherically
symmetric vacuum solution is static and uniquely given by the
Schwarzschild geometry, which exhibits a curvature singularity at
$r=0$. In pseudo-complex general relativity, the modified field
equations \eqref{eq:pceinstein} invalidate the standard form of
Birkhoff's theorem. Static solutions can still be constructed, but their
properties follow from the pseudo-complex constraint structure rather
than from a general uniqueness result.

The static, spherically symmetric metric considered below is the
pseudo-complex Schwarzschild-type solution obtained from the
pseudo-complex Einstein equations discussed in
Sec.~\ref{ssec:field_equations}. Its derivation and subsequent
extensions are given in
Refs.~\cite{HessGreiner2009,Caspar2012}.

For the physical metric sector, we define the lapse function by
\begin{equation}
A(r)\equiv e^{\nu(r)},
\label{eq:lapse-definition}
\end{equation}
the metric takes the form
\begin{equation}
ds^2 = A(r)\,dt^2 - \frac{1}{A(r)} \left( 1-\frac{a_0^2}{A(r)^2}
\right)^{-1} dr^2 -r^2d\Omega^2.
\label{eq:ss_metric}
\end{equation}
Here $a_0$ is the invariant acceleration scale introduced in
Sec.~\ref{sec:2}. Spherical symmetry restricts the auxiliary
sector $f_{\mu\nu}$ to depend only on $r$, and its components
are determined by the requirement that both idempotent sectors
satisfy the field equations simultaneously.

In contrast to the Schwarzschild solution, the relation
$g_{rr}=g_{tt}^{-1}$ is modified, reflecting the underlying
pseudo-complex consistency conditions.
The auxiliary radial function $B(r)$ introduced below should
not be confused with the lapse function $A(r)$.

\subsection{Origin of the lapse-gap condition}

The lower bound on the lapse function should not be viewed as an
independent regularity assumption imposed after the fact. It is the local
form, in the static spherically symmetric sector, of the admissibility
condition introduced by the pseudo-complex extension.

To see this, note that the radial metric coefficient in Eq.~\eqref{eq:ss_metric}
contains the factor
\begin{equation}
\Delta_{\rm pc}(r)
\equiv
1-\frac{a_0^2}{e^{2\nu(r)}} .
\label{eq:pc_gap_factor}
\end{equation}
For the physical metric to remain Lorentzian and nondegenerate in the
radial sector, this factor must not change sign or vanish in the region
under consideration. Therefore admissible configurations satisfy
\begin{equation}
\Delta_{\rm pc}(r)>0,
\label{eq:pc_admissibility}
\end{equation}
which is equivalent to
\begin{equation}
e^{\nu(r)} > a_0 .
\label{eq:lapse_gap_origin}
\end{equation}
Thus the lapse-gap condition follows from the requirement that the
pseudo-complex corrected radial metric remain a regular Lorentzian
metric.

In this sense, the pseudo-complex scale $a_0$ acts as a geometric
exclusion scale: configurations for which the lapse approaches zero in
the Schwarzschild manner are not part of the admissible solution space.
The singular Schwarzschild behavior is therefore removed not by adding
a new matter source, but by restricting the metric configurations allowed
by the pseudo-complex geometry.

For the local curvature estimates below we use the slightly stronger
uniform version of this condition,
\begin{equation}
e^{2\nu(r)}-a_0^2 \ge \delta a_0^2 ,
\qquad \delta>0,
\label{eq:uniform_lapse_gap}
\end{equation}
which ensures that the metric remains bounded away from degeneracy in
a neighborhood of the origin. This uniform gap is the condition needed
to obtain explicit bounds on curvature invariants.  

\subsection{Curvature regularization near the center}

We now state the curvature-regularization result in a form that makes
the underlying assumptions explicit. The pseudo-complex constraint
structure restricts the admissible class of metric functions, and within
this class the resulting conditions are sufficient to ensure bounded
curvature in a neighborhood of the center.

Consider a physical metric of the form
\begin{equation}
ds^2 = A(r)\,dt^2 - B(r)\,dr^2 -r^2 d\Omega^2 , 
\label{eq:ss_metric_A}
\end{equation}
where
\begin{equation}
A(r) = e^{\nu(r)}
\end{equation}
and
\begin{equation}
B(r) = \frac{1}{A(r)} \left(1-\frac{a_0^2}{A(r)^2}\right)^{-1}.
\end{equation}
For the metric to describe a regular geometry in a neighborhood of the origin,
the lapse function must satisfy standard regularity conditions. In particular,
for $0 \le r < \epsilon$, we require
\begin{equation}
A(r)^2-a_0^2 \ge \delta a_0^2,
\qquad \delta>0,
\label{eq:lapse_gap_prop}
\end{equation}
which is precisely the nondegeneracy condition imposed by the
pseudo-complex structure. Moreover, $A(r)$ is twice continuously
differentiable with
\begin{equation}
|A'(r)|\le B_1, \qquad |A''(r)|\le B_2 ,
\label{eq:A_bounds_prop}
\end{equation}
where a prime denotes differentiation with respect to $r$.

We now examine the implications of the regularity conditions at the
areal-radius origin,
\begin{equation}
B(0)=1, \qquad B'(0)=0 .
\label{eq:center_regularity}
\end{equation}
In the pcGR framework, these standard geometric consistency conditions
acquire additional significance because they impose compatibility
constraints on the lapse function through the deformed relation
between $A(r)$ and $B(r)$.  The condition $B(0)=1$ implies
\begin{equation}
\frac{1}{A(0)} \left( 1-\frac{a_0^2}{A(0)^2} \right)^{-1} =1,
\end{equation}
which relates the central value of the lapse to the
pseudo-complex scale,
\begin{equation}
A(0)^2-a_0^2 = A(0)^2 \left( 1-\frac{a_0^2}{A(0)^2} \right) =1.
\end{equation}
In the limit $a_0\to0$, this reduces to
$A(0)=1$, recovering the standard GR regularity
condition.

Similarly, the condition $B'(0)=0$ constrains $A'(0)$ through the
relation between $B(r)$ and $A(r)$. Differentiating $B(r)$ and
evaluating the result at $r=0$ shows that $B'(0)=0$ implies $A'(0)=0$
(or, more generally, the corresponding compatibility relation),
consistent with local flatness at the origin.  In classical GR, the
Schwarzschild solution is incompatible with these conditions because
$A(r)\to0$ as $r\to0$.  In pcGR, with $A(0)>a_0$, they are naturally
compatible with the deformed metric structure.

Under these conditions, curvature invariants constructed from
contractions of the Riemann tensor remain finite at $r=0$. These
conditions are sufficient for curvature regularity within the class of
static, spherically symmetric configurations considered here, but are
not claimed to be necessary in general. In particular, the Kretschmann scalar
\begin{equation}
K(r)\equiv R_{\mu\nu\rho\sigma}R^{\mu\nu\rho\sigma}
\end{equation}
satisfies
\begin{equation}
K(r)\le C(\delta,B_1,B_2)\,a_0^{-4},
\qquad r\to 0,
\label{eq:K_bound_prop}
\end{equation}
where $C(\delta,B_1,B_2)$ is finite and dimensionless.

This follows from the structure of the metric coefficients. The gap
condition \eqref{eq:lapse_gap_prop} keeps $A(r)^2-a_0^2$ bounded away
from zero, so the radial metric component and its first two derivatives
remain finite whenever $A$, $A'$, and $A''$ are bounded. Since the
Riemann tensor depends only on the metric and its first and second
derivatives, no divergence can arise from these terms.

The only additional possible singularity comes from the angular part of
the curvature, which contains contributions of the form
\begin{equation}
\frac{1}{r^4}\left(1-\frac{1}{B(r)}\right)^2 .
\end{equation}
The regular-center conditions \eqref{eq:center_regularity} imply
\begin{equation}
\frac{1}{B(r)} = 1+O(r^2),
\end{equation}
and therefore this angular contribution remains finite as $r\to0$.
Thus the curvature is bounded, with the local scale set by $a_0$.
The bound may be written in the form \eqref{eq:K_bound_prop} after
restoring the physical invariant acceleration scale associated with the
dimensionless parameter $a_0$.

Equivalently, writing $A(r)=e^{\nu(r)}$, the gap condition may be expressed as
\begin{equation}
e^{\nu(r)} \ge (1+\delta)^{1/2} a_0,
\qquad \delta > 0,
\end{equation}
with bounded derivatives of $\nu(r)$ in a neighborhood of the origin.


\subsection{General-relativistic limit and weak-field expansion}

It is useful to make explicit how the pseudo-complex solution reduces
to the Schwarzschild geometry in the appropriate limit and how the
modifications enter in the weak-field regime.
Consider the regime in which the pseudo-complex correction is small,
\begin{equation}
\frac{a_0^2}{e^{2\nu(r)}} \ll 1 .
\end{equation}
In this limit, the radial metric component becomes
\begin{equation}
\begin{split}
g_{rr} &= -e^{-\nu(r)} \left( 1-\frac{a_0^2}{e^{2\nu(r)}} \right)^{-1}
\\ &= -e^{-\nu(r)} \left[ 1+\frac{a_0^2}{e^{2\nu(r)}} +\mathcal O
  \!\left( \frac{a_0^4}{e^{4\nu(r)}} \right) \right].
\end{split}
\end{equation}
To leading order,
\begin{equation}
g_{rr}\simeq -e^{-\nu(r)},
\end{equation}
so that the standard Schwarzschild relation
\(g_{rr}=g_{tt}^{-1}\)
is recovered. The pseudo-complex corrections are therefore suppressed,
and the solution reduces continuously to general relativity.

The weak-field reduction is governed by the dimensionless expansion
parameter
\[
\epsilon\equiv\frac{a_0^2}{e^{2\nu(r)}}.
\]
Expanding the deformed lapse function,
\[
A(r) = e^{\nu(r)} \left( 1-\frac{a_0^2}{e^{2\nu(r)}} \right)^{1/2},
\]
for \(\epsilon\ll1\) gives
\[
A(r) = e^{\nu(r)} - \frac{a_0^2}{2e^{\nu(r)}} + \mathcal
O(\epsilon^2).
\]
Thus, \(\epsilon\) is the parameter controlling the recovery of GR, and
the pseudo-complex corrections become significant only in the
high-curvature regime.

In the Newtonian limit,
\begin{equation}
e^{\nu(r)}=1+\Phi(r), \qquad |\Phi(r)|\ll1,
\end{equation}
one finds
\begin{equation}
g_{rr} \simeq -1+\Phi-a_0^2 +\mathcal O(\Phi a_0^2,a_0^4).
\end{equation}
The additional term proportional to \(a_0^2\) should not be interpreted
as an independent Newtonian potential. Physical post-Newtonian
constraints apply to coordinate-invariant combinations of the metric
functions, and the pseudo-complex corrections remain suppressed
throughout the regime
\(a_0^2/e^{2\nu(r)}\ll1\).

Accordingly, pcGR reduces smoothly to general relativity in the
weak-field limit while producing observable deviations only in the
strong-field regime, where the pseudo-complex corrections become
appreciable. This behavior is consistent with the established pcGR
literature
\cite{HessGreiner2009,Caspar2012,Schonenbach2014,Schonenbach2016,NielsenBirnholtz2017,Maimon2025}.


\subsection{Near-horizon structure and observable scale}

The pseudo-complex modification alters the radial metric component in
regions where the lapse approaches its lower admissible value,
suggesting a modification of the near-horizon geometry of compact
objects.

For comparison, the Schwarzschild lapse is
\begin{equation}
A_{\rm Schw}(r)=1-\frac{2GM}{r},
\end{equation}
with a horizon at \(r_s=2GM\), where \(A_{\rm Schw}(r_s)=0\).  In the
pseudo-complex framework, admissible configurations satisfy
\begin{equation}
A(r)\ge a_0,
\end{equation}
so the lapse remains bounded away from zero within the regular branch
of the solution.

Estimating the onset of pseudo-complex effects by applying the
admissibility condition to the Schwarzschild lapse gives
\begin{equation}
A_{\rm Schw}(r_{\rm min})\simeq a_0,
\end{equation}
or
\begin{equation}
r_{\rm min}\simeq\frac{2GM}{1-a_0},
\end{equation}
for the dimensionless parameter \(a_0\ll1\). The corresponding
relative shift is
\begin{equation}
\frac{r_{\rm min}-r_s}{r_s}\simeq a_0.
\end{equation}

This estimate should not be interpreted as the exact location of a
modified horizon. Rather, it identifies the characteristic scale at
which deviations from the Schwarzschild geometry become significant.
The complete near-horizon structure must be obtained by solving the
pseudo-complex field equations.

Such deviations may affect strong-field observables, including
quasi-normal mode spectra, gravitational-wave ringdown, and
photon-sphere properties. To leading order, one expects
\begin{equation}
\frac{\Delta\omega}{\omega}\sim a_0,
\end{equation}
and similarly
\begin{equation}
\frac{\Delta r_{\rm ph}}{r_{\rm ph}}\sim a_0.
\end{equation}
A detailed phenomenological analysis lies beyond the scope of the
present work, but these observables provide potential probes of the
pseudo-complex scale.


\section{Cosmological Implications}\label{sec:cosmo}

Cosmology provides a natural setting to analyze the dynamical consequences
of the pseudo-complex constraint structure. In a homogeneous and isotropic
universe, the physical metric takes the standard
Friedmann--Lema\^itre--Robertson--Walker form,
\begin{equation}
ds^2 = dt^2 - a(t)^2 \left( \frac{dr^2}{1 - k r^2} + r^2 d\Omega^2 \right),
\end{equation}
where $k = 0, \pm 1$.

Spatial isotropy restricts the auxiliary sector to the form
\begin{equation}
f_{tt}(t), 
\qquad 
f_{rr}(t) \propto \frac{a(t)^2}{1 - k r^2},
\end{equation}
with angular components fixed by symmetry and no allowed vector or mixed
terms. The pseudo-complex extension therefore preserves the symmetry of
the cosmological background while modifying the relation between the
metric components.

Projecting the pseudo-complex Einstein equations onto the idempotent
sectors yields two coupled Friedmann systems for the effective metrics
$g^{(\pm)}_{\mu\nu} = g_{\mu\nu} \pm f_{\mu\nu}$,
\begin{equation}
H_\pm^2 = \frac{\kappa}{3} \rho^{\text{eff}}_\pm, \qquad \dot{H}_\pm =
-\frac{\kappa}{2} \left( \rho^{\text{eff}}_\pm + p^{\text{eff}}_\pm
\right),
\end{equation}
where $H = \dot{a}/a$.  In the physical sector, the Friedmann equation
can be written as
\begin{equation}
H^2 = \frac{\kappa}{3} \rho_{\text{matter}} + \frac{\kappa}{3}
\rho_{\text{pc}}(t),
\label{eq:46}
\end{equation}
where $\rho_{\text{pc}}(t)$ denotes an effective energy density induced
by the pseudo-complex geometry.

We now present a formal reduction of the pseudo-complex
cosmological field equations to an effective evolution equation
for $\rho_{\rm pc}(t)$. The consistency of the two idempotent
sectors implies that $f_{\mu\nu}(t)$ is determined algebraically
in terms of $a(t)$ and its derivatives. Under the assumption
that the resulting differential operators admit the required
inverse operators and that the auxiliary functions may be
eliminated consistently, the system reduces to the differential
equation
\begin{equation}
\dot{\rho}_{\rm pc} +3H(\rho_{\rm pc}+p_{\rm pc}) =\Gamma_{\rm pc}(t),
\label{eq:closed_evolution}
\end{equation}
where, under the assumptions stated above,
$p_{\rm pc}$ and $\Gamma_{\rm pc}$ are determined
by $\rho_{\rm pc}$ and its derivatives
 (or, equivalently, in terms
of $a(t)$ and its derivatives), so that no arbitrary auxiliary
functions remain.
The formal elimination procedure leading to these expressions is
presented in \ref{sec:appendixB}, where we show how the system can be reduced
to a single second-order differential equation for $\rho_{\rm pc}(t)$.
Specifically, the formal elimination of the auxiliary functions $f_0$
and $f_2$ leads to an effective second-order evolution equation for
$\rho_{\rm pc}(t)$ of the form
\begin{eqnarray}
\ddot{\rho}_{\rm pc} + \mathcal{A}(\rho_{\rm pc}, \dot{\rho}_{\rm pc},
a, \dot{a}) \dot{\rho}_{\rm pc} + \mathcal{B}(\rho_{\rm pc}, a,
\dot{a}) \rho_{\rm pc} \nonumber \\
= \mathcal{S}(\rho_{\rm pc}, \dot{\rho}_{\rm
  pc}, a, \dot{a}),
\label{eq:closed_explicit}
\end{eqnarray}
where $\mathcal{A}, \mathcal{B}, \mathcal{S}$ are explicitly defined
in \ref{sec:appendixB}. Within the assumptions of
\ref{sec:appendixB}, this equation, together with the
Friedmann equation for $a(t)$, provides a formal representation of the
reduced cosmological dynamics of the pseudo-complex vacuum component.

In the general case, Eq.~\eqref{eq:closed_evolution} does not admit
simple analytic solutions. However, its structure reveals that
$\rho_{\text{pc}}(t)$ is generically time dependent, with its evolution
determined by the geometry. The equation admits solutions that decay
as the universe expands, as well as solutions that exhibit transient
behavior, depending on the initial conditions for $a(t)$. A detailed
analysis of these solutions will be presented elsewhere.

Dimensional analysis fixes the characteristic scale of this contribution.
Let $a_{0,\rm phys}$ denote the physical inverse-length scale associated
with the pseudo-complex extension. The induced curvature corrections are
then of order $a_{0,\rm phys}^2$. Using the relation $H^2\sim\kappa\rho$,
this implies
\begin{equation}
\rho_{\text{pc}} \sim \frac{a_{0,\rm phys}^2}{\kappa},
\end{equation}
so that the physical invariant acceleration scale sets the magnitude of
the effective vacuum energy.

In general, $\rho_{\text{pc}}(t)$ is time dependent, with behavior that
can include decaying or transient evolution depending
on the dynamics of the auxiliary sector as determined by
Eq.~\eqref{eq:closed_evolution}. This effective contribution
arises without introducing scalar fields or potentials and instead
reflects the constraint structure of the pseudo-complex geometry.
The underlying mechanism differs from standard field-theoretic models, where
vacuum evolution is typically governed by scalar potentials.

\subsection{Dynamical vacuum structure: asymptotic regimes and
  physical interpretation}

The formally reduced equation for \( \rho_{\rm pc}(t) \),
Eq.~\eqref{eq:closed_explicit}, reveals a rich dynamical
structure. The function \(\mathcal{A}\) acts as an effective friction
term, damping or amplifying \(\rho_{\rm pc}\) depending on the
expansion history.  The term \( \mathcal{B} \) plays the role of a
time-dependent effective mass, coupling the vacuum component to the
geometry. The source \( \mathcal{S} \) represents the driving force
from matter and curvature, ensuring that \( \rho_{\rm pc} \) is not a
static quantity but responds to the cosmic evolution.  To understand
the physical implications, we analyze the asymptotic behavior of \(
\rho_{\rm pc}(t) \) in two limiting regimes.

\paragraph{Early-universe limit \( a \to 0 \).} 
In the early universe, the Friedmann equation gives \( H \gg 1 \) and
the curvature term \( k/a^2 \ll H^2 \). 
The operators $\mathcal{D}$ and $\mathcal{E}$, defined in
Eqs.~\eqref{eq:D_operator} and \eqref{eq:E_operator},
reduce to
\begin{equation}
\mathcal{D} \simeq H \frac{d}{dt}, \qquad 
\mathcal{E} \simeq \frac{d^2}{dt^2} + 2H \frac{d}{dt}.
\label{eq:D_E_operators}
\end{equation}
Substituting these expressions into the formally reduced equation gives,
at leading order,
\begin{equation}
\ddot{\rho}_{\rm pc}+3H\dot{\rho}_{\rm pc}\simeq0,
\end{equation}
and hence
\begin{equation}
\dot{\rho}_{\rm pc}=C\,a^{-3},
\end{equation}
where \(C\) is an integration constant. For a radiation-dominated
background, \(a(t)\propto t^{1/2}\), this gives
\begin{equation}
\rho_{\rm pc}(t)=\rho_0+C_{\rm r}t^{-1/2} =\rho_0+\widetilde C_{\rm
  r}a^{-1},
\end{equation}
whereas for a matter-dominated background,
\(a(t)\propto t^{2/3}\),
\begin{equation}
\rho_{\rm pc}(t)=\rho_0+C_{\rm m}t^{-1} =\rho_0+\widetilde C_{\rm
  m}a^{-3/2}.
\end{equation}
Thus the leading-order equation admits both a constant solution
(\(C=0\)) and branches that grow toward the past. Finiteness of
\(\rho_{\rm pc}\) as \(a\to0\) therefore requires the constant branch
or additional conditions supplied by the full reduced equations.
The leading-order approximation alone does not establish generic
early-time regularity.

\paragraph{Late-time limit \( a \to \infty \).}
In the late universe, assuming matter domination, \( H \sim t^{-1} \)
and the curvature term \( k/a^2 \) becomes negligible. The operators
reduce to algebraic forms and the formally reduced equation admits
asymptotic solutions that approach an effective constant vacuum energy
under the corresponding late-time assumptions:
\[
\rho_{\rm pc} \to \rho_\infty \qquad \text{as } a \to \infty.
\]
This corresponds to an effective cosmological constant. The value of
\( \rho_\infty \) is set by the same geometric scale that regularizes
black-hole curvature,
\[
\rho_\infty \sim \frac{a_{0,\rm phys}^2}{\kappa}.
\]
Thus, the same parameter \( a_0 \) that enforces the gap \( e^{\nu(r)}
> a_0 \) in spherical symmetry also fixes the characteristic magnitude
of the vacuum energy. This provides a unified geometric origin for
both black-hole regularity and cosmic acceleration.

\paragraph{Effective equation of state and future work.}
The effective equation of state
\[
p_{\rm pc} = \mathcal{P}(\rho_{\rm pc}, \dot{\rho}_{\rm pc}, a, \dot{a})
\]
determines whether the vacuum component behaves as a cosmological
constant (\( w = -1 \)), quintessence (\( -1 < w < -1/3 \)), or a more
exotic form. Its explicit form requires the full nonlinear expansion
of \( \mathcal{A}, \mathcal{B}, \mathcal{S} \) in terms of \(
\rho_{\rm pc} \) and \( a \), which is left for future
work. Nevertheless, the asymptotic analysis above already reveals that
pcGR naturally accommodates a dark-energy component with a geometric
origin, with the scale controlled by the invariant acceleration
parameter \( a_0 \).


\section{Comparison with Quantum-Gravity Motivated Constraints}
\label{sec:swampland}

Having established the main technical results of the pseudo-complex
framework in Secs.~\ref{sec:pcgr_solutions} and \ref{sec:cosmo}, we
now turn to a structural comparison with constraints that have been
proposed in quantum-gravity settings.  The purpose of this comparison
is not to claim that pcGR satisfies or derives these conjectures, but
rather to identify which features commonly associated with
quantum-gravity consistency can already arise at the level of
classical geometry.

A recurring theme in quantum gravity is that physically admissible
configurations are subject to intrinsic limitations. These include bounds on
curvature scales, restrictions on global symmetries, and constraints on vacuum
structure. Such conditions typically arise within effective field theory as
consistency requirements motivated by black-hole physics, string
compactifications, holography, and anomaly considerations
\cite{Vafa2005,OoguriVafa2007,Palti2019}.

From the perspective of the present work, the central point is that
pseudo-complex general relativity imposes analogous restrictions through purely
geometric consistency conditions, modifying the geometric structure of
spacetime itself rather than introducing new dynamical degrees of freedom.
This motivates a comparison at the level of structural features rather than
at the level of microscopic realizations.

The Swampland program \cite{Palti2019} provides a set of conjectured
criteria that any consistent theory of quantum gravity must satisfy,
representing one of the most prominent frameworks for
quantum-gravity-inspired constraints. Among these, the Distance
Conjecture \cite{Vafa2005,OoguriVafa2007} asserts that
infinite-distance limits in moduli space are accompanied by a tower of
light states, signaling a breakdown of the effective field
theory. This conjecture constrains the field-space geometry of any UV
completion and has direct implications for models of inflation and
dark energy. The de Sitter conjectures \cite{OoguriVafa2007} further
restrict the scalar potential, forbidding stable de Sitter vacua and
requiring \( |\nabla V| \ge c V \) or \( \min(\nabla_i\nabla_j V) \le
-c' V \). These conditions are particularly relevant for cosmology, as
they challenge the standard paradigm of slow-roll inflation and
dark-energy domination. The cobordism conjecture \cite{vanBeest2022}
and the absence of global symmetries \cite{OoguriPaltiShiuVafa2019}
impose additional topological and algebraic constraints on the gauge
structure of the theory. While the pcGR framework is not a full UV
completion, its pseudo-complex algebraic structure may offer a
geometric setting that naturally accommodates or bypasses some of
these constraints, particularly through the introduction of a minimum
length scale encoded in \( a_0 \), which regularizes the central
singularity and modifies the strong-field geometry.  The connection
between pcGR and Hořava-Lifshitz gravity
\cite{HessZenVasconcellosHadjimichef2025} further suggests that the
extended coordinate formulation of gravity may provide a bridge
between phenomenological regular black-hole solutions and the deeper
principles of quantum gravity. These connections, while speculative,
point toward fruitful avenues for future investigation, where the
interplay between pseudo-complex geometry, modified gravity, and
quantum-gravity-inspired con- \break straints may lead to a more complete
understanding of the ultraviolet structure of gravity.

\subsection{Geometric Consequences of the Pseudo-Complex Constraint Structure}

The pseudo-complex extension of spacetime imposes algebraic consistency
conditions linking the two idempotent sectors of the theory. These
constraints restrict the class of admissible metric configurations and
lead to a set of well-defined geometric consequences. In this section we
identify the principal effects of this constraint structure and analyze
how they manifest in both local and cosmological settings.

\subsubsection{Bounded curvature and high-curvature regularity}

Among the consequences of the pseudo-complex constraint structure is the
regularization of curvature in high-symmetry configurations, as derived in
Sec.~\ref{sec:pcgr_solutions}. The
invariant acceleration scale $a_0$ imposes a lower bound on the lapse
function, which prevents the vanishing behavior responsible for
curvature divergences in solutions such as the Schwarzschild geometry.

As a result, curvature invariants remain finite in regions where general
relativity would predict singular behavior. This regularization mechanism
is entirely geometric: it does not rely on additional matter sources,
modified dynamics, or quantum corrections, but follows directly from
restrictions on admissible spacetime configurations, together
with the consistency conditions at the areal-radius origin derived in
Sec.~\ref{sec:pcgr_solutions}.

\subsubsection{Intrinsic cutoff scale}

The pseudo-complex framework introduces an intrinsic geometric scale
through the invariant acceleration parameter $a_0$. This scale controls
the behavior of the metric in high-curvature regimes and effectively
limits access to arbitrarily large curvature.

Unlike ultraviolet cutoffs arising from new dynamical degrees of freedom,
the scale $a_0$ enters through the constraint structure of the geometry
itself. The resulting cutoff is therefore kinematic in origin, reflecting
a restriction on the allowed spacetime configurations rather than a
breakdown of the effective description.

\subsubsection{Absence of additional conserved charges}

The auxiliary sector \( f_{\mu\nu} \) is determined algebraically and does not
introduce independent dynamical fields. Consequently, the extended geometric
structure does not give rise to new conserved charges associated with
additional symmetries. This reflects a general feature of the pseudo-complex
construction: all additional structure is encoded in constraints rather than in
new symmetries or new degrees of freedom.

\subsubsection{Dynamical effective vacuum component}

In cosmological settings, the pseudo-complex geometry generates a
dynamical vacuum component in the Friedmann equations, as shown in
Sec.~\ref{sec:cosmo}. This term arises from the geometric structure of
the theory rather than from an independent scalar field or potential.

The resulting vacuum energy is therefore intrinsically dynamical, with
its evolution determined by the constraint relations governing
$f_{\mu\nu}$, as encoded in Eq.~\eqref{eq:closed_evolution}
and formally represented by Eq.~\eqref{eq:closed_explicit}.
This provides a purely geometric mechanism for generating
time-dependent vacuum behavior without introducing additional
fields or modifying the propagating degrees of freedom.

\subsubsection{Limitations of the geometric framework}

While the pseudo-complex constraint structure leads to several
nontrivial geometric effects, its scope is restricted to modifications of
spacetime structure at the classical level. In particular, it does not
introduce new microscopic degrees of freedom or modify the spectrum of
excitations.

As a consequence, phenomena that depend explicitly on quantum spectra,
such as the emergence of towers of light states, lie outside the domain
of the pseudo-complex framework. This limitation delineates the boundary
between geometric constraints and genuinely quantum-gravitational
effects.

\begin{figure*}[htbp]
  \centering
  \resizebox{0.9\linewidth}{!}{%
    \begin{tikzpicture}[
      node distance=10mm and 18mm,
      box/.style={
        draw,
        rounded corners,
        align=center,
        inner sep=6pt,
        minimum height=1cm
      },
      titlebox/.style={
        box,
        text width=6.2cm
      },
      midbox/.style={
        box,
        text width=4.8cm
      },
      listbox/.style={
        draw,
        rounded corners,
        align=left,
        inner sep=6pt,
        text width=5.2cm
      },
      bottombox/.style={
        box,
        text width=10.8cm
      },
      arrow/.style={
        -{Latex[length=2.5mm]},
        semithick
      }
    ]

    \node[titlebox] (top) {\textbf{Swampland-inspired\\ structural features}};

    \node[midbox, below left=of top] (leftmid) {\textbf{Geometric features\\ realized in pcGR}};
    \node[midbox, below right=of top] (rightmid) {\textbf{Features without direct\\ geometric analogs in pcGR}};

    \node[listbox, below=of leftmid] (leftlist) {
    $\bullet$ Bounded curvature\\[2pt]
    $\bullet$ Intrinsic cutoff scale\\[2pt]
    $\bullet$ Dynamical vacuum behavior\\[2pt]
    $\bullet$ Absence of exact global symmetries
    };

    \node[listbox, below=of rightmid] (rightlist) {
    $\bullet$ Infinite towers of light states\\[2pt]
    $\bullet$ Distance conjecture tower behavior\\[2pt]
    $\bullet$ Other explicitly spectral quantum constraints
    };

    \node[bottombox, below=18mm of $(leftlist.south)!0.5!(rightlist.south)$] (bottom) {
    \textbf{Main conclusion:} pcGR exhibits a subset of Swampland-like features
    through consistency conditions, while others remain intrinsically
    tied to quantum degrees of freedom.
    };

    \draw[arrow] (top) -- (leftmid);
    \draw[arrow] (top) -- (rightmid);
    \draw[arrow] (leftmid) -- (leftlist);
    \draw[arrow] (rightmid) -- (rightlist);
    \draw[arrow] (leftlist.south) |- (bottom.west);
    \draw[arrow] (rightlist.south) |- (bottom.east);

    \end{tikzpicture}
  }

  \caption{Schematic classification of Swampland-inspired features in
    pseudo-complex general relativity (pcGR). Some features arise from
    geometric constraints, while others require quantum degrees of
    freedom.}
  \label{fig:pcgr_swampland_classification}

\end{figure*}
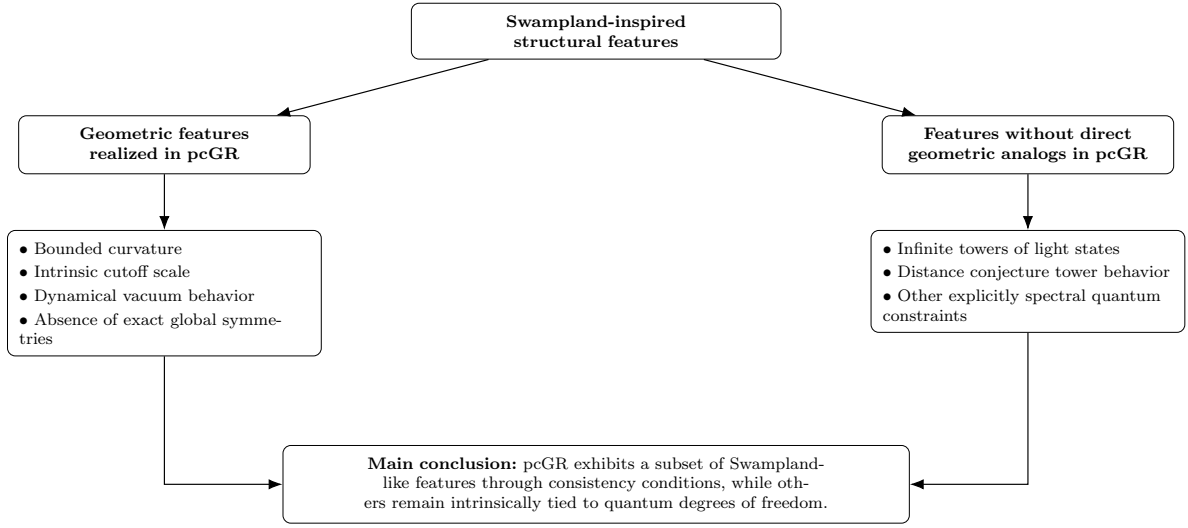

Taken together, these results establish pseudo-complex general relativity
as a constrained geometric framework in which curvature regularization,
intrinsic scales, and dynamical vacuum behavior arise directly from the
structure of spacetime. Figure~\ref{fig:pcgr_swampland_classification}
summarizes this classification.

\section{Outlook on Possible Ultraviolet Structures}
\label{sec:spec.outlook}

The results of the present work establish pcGR as a consistent
geometric framework for regular black-hole solutions and nonsingular
cosmologies \cite{HessGreiner2009,Caspar2012}. The lapse-gap condition
\( e^{\nu(r)} > a_0 \), together with the regularity conditions \(
B(0)=1 \) and \( B'(0)=0 \) derived in Sec.~\ref{sec:pcgr_solutions},
provides a controlled mechanism for eliminating the Schwarzschild-type
curvature divergence while recovering GR in the weak-field limit
\cite{HessGreiner2009}. The formally reduced cosmological  equation,
Eq.~\eqref{eq:closed_explicit} from Sec.~\ref{sec:cosmo}, governs the
effective dynamics of the 
pseudo-complex energy density \( \rho_{\rm pc}(t) \),
replacing the phenomenological dark-energy component with a geometric
term arising from the pseudo-complex structure of the field equations
\cite{HessMaghlaouiGreiner2010,Maghlaoui2026}.

The simultaneity condition, namely that the pseudo-complex
Einstein equations must hold independently in both idempotent sectors
\(e_+\) and \(e_-\),  follows from the algebraic structure of the
pseudo-complex numbers and is derived in the foundational work
\cite{HessGreiner2009}. This simultaneity condition is not merely a
formal requirement; it enters explicitly in the derivation of the
curvature bounds in \ref{sec:appendixA}, where the independence of the two
idempotent sectors ensures that the denominator \( A(r)^2 - a_0^2 \)
controls the entire invariant spectrum (see Eqs.~\eqref{eq:A.5}--\eqref{eq:A.7} and the
surrounding discussion).

The pseudo-complex extension of spacetime introduced in this work is
formulated entirely at the level of classical geometry. It does not rely
on an underlying microscopic theory, and no claim is made here regarding
a derivation from a specific ultraviolet (UV) completion. Nevertheless,
it is useful to examine whether the geometric structures identified above
admit connections to frameworks that arise in quantum gravity.

The spherically symmetric solutions derived in
Sec.~\ref{sec:pcgr_solutions}, the formally reduced cosmological
evolution equation from
Sec.~\ref{sec:cosmo} [Eq.~\eqref{eq:closed_explicit}], and the geometric constraint
structure discussed in Sec.~\ref{sec:swampland} provide the concrete results that
motivate the following speculative connections.

\begin{table*}
  \caption{\label{tab:qg_comparison} Comparison of pseudo-complex
    general relativity with representative approaches to quantum
    gravity that address bounded curvature, minimal length, and
    singularity resolution. The entries highlight structural features
    and qualitative mechanisms. Here, EFT denotes
      effective field theory. The comparison does not imply a
    technical equivalence, derivation, or embedding of pseudo-complex
    general relativity within these frameworks.}  \centering
\begin{tabular}{@{}lllll@{}}
\toprule
\textbf{Aspect} &
\textbf{pcGR} &
\textbf{Loop Quantum Gravity} &
\textbf{Wheeler--DeWitt/} &
\textbf{String Theory} \\
\textbf{} &
\textbf{} &
\textbf{} &
\textbf{Quantum Cosmology} &
\textbf{}\\ 
\midrule 

\parbox[t]{0.14\textwidth}{\raggedright  Primary\\ modification} &
\parbox[t]{0.18\textwidth}{\raggedright  Classical geometric extension\\(pseudo-complex spacetime; invariant acceleration scale)} &
\parbox[t]{0.18\textwidth}{\raggedright  Quantization of geometry\\(holonomies/fluxes; discrete spectra)} &
\parbox[t]{0.20\textwidth}{\raggedright  Canonical quantization of gravity\\(wave function of the universe; minisuperspace)} &
\parbox[t]{0.18\textwidth}{\raggedright  Microscopic UV completion\\(extended objects; extra dimensions; dualities)} \\
\noalign{\vskip 5pt}

\parbox[t]{0.14\textwidth}{\raggedright  Spacetime description} &
\parbox[t]{0.18\textwidth}{\raggedright  Smooth manifold with constrained short-distance structure} &
\parbox[t]{0.18\textwidth}{\raggedright  Fundamentally discrete geometry; continuum as an effective limit} &
\parbox[t]{0.20\textwidth}{\raggedright  Quantum state on superspace; classical spacetime emerges semiclassically} &
\parbox[t]{0.18\textwidth}{\raggedright  Typically higher-dimensional; effective 4D spacetime after compactification} \\
\noalign{\vskip 5pt}

\parbox[t]{0.14\textwidth}{\raggedright  Minimal-length/\\ UV mechanism} &
\parbox[t]{0.18\textwidth}{\raggedright  Geometric cutoff via maximal acceleration / bounded curvature} &
\parbox[t]{0.18\textwidth}{\raggedright  Area and volume gaps; polymerized effective dynamics} &
\parbox[t]{0.20\textwidth}{\raggedright  Quantum boundary conditions suppress classical singularities} &
\parbox[t]{0.18\textwidth}{\raggedright  Fundamental string length and T-duality limit spatial resolution} \\
\noalign{\vskip 8pt}

\parbox[t]{0.14\textwidth}{\raggedright  Singularity resolution} &
\parbox[t]{0.18\textwidth}{\raggedright  Regular cores in symmetric solutions\\(geometric regularization)} &
\parbox[t]{0.18\textwidth}{\raggedright  Bounces or regular interiors\\(quantum geometry)} &
\parbox[t]{0.20\textwidth}{\raggedright  Singularity avoidance via quantum dynamics or boundary conditions} &
\parbox[t]{0.18\textwidth}{\raggedright  Resolution via extended objects, dual descriptions, or microstate structure} \\
\noalign{\vskip 8pt}

\parbox[t]{0.14\textwidth}{\raggedright  New microscopic\\ degrees of freedom\\ required?} &
\parbox[t]{0.18\textwidth}{\raggedright  No\\(auxiliary sector fixed algebraically)} &
\parbox[t]{0.18\textwidth}{\raggedright  Yes\\(quantum geometric degrees of freedom)} &
\parbox[t]{0.20\textwidth}{\raggedright  Yes\\(quantum gravitational degrees of freedom)} &
\parbox[t]{0.18\textwidth}{\raggedright  Yes\\(strings, branes, and associated spectra)} \\
\noalign{\vskip 8pt}

\parbox[t]{0.14\textwidth}{\raggedright  Relation to\\ EFT constraints} &
\parbox[t]{0.18\textwidth}{\raggedright  Not formulated as an EFT framework; constraints arise geometrically} &
\parbox[t]{0.18\textwidth}{\raggedright  Not formulated as an EFT diagnostic} &
\parbox[t]{0.20\textwidth}{\raggedright  Not formulated as an EFT diagnostic} &
\parbox[t]{0.18\textwidth}{\raggedright  Source of consistency conditions in low-energy EFTs} \\
\bottomrule
\end{tabular}
\end{table*}

One possible point of contact is provided by para-complex geometry,
which appears in $\mathcal{N}=2$ supergravity and special K{\"a}hler
geometry \cite{Cortes2004,deWitVanProeyen1992,Freed1999}. Para-complex
structures satisfy $J^2 = +1$ and admit an idempotent decomposition
similar to that used in the pseudo-complex construction. In supergravity,
these structures govern scalar-field couplings, whereas in the present
framework they act directly on spacetime. This analogy suggests that the
pseudo-complex extension may be viewed as a geometric implementation of
structures that appear in more general settings.

A related observation arises in doubled field theory and generalized
geometry, where the coordinate space is extended to include dual
coordinates \cite{HullZwiebach2009,HohmHullZwiebach2010,Hitchin2003}.
The generalized metric in such formulations contains multiple symmetric
components, reminiscent of the decomposition
\begin{equation}
G_{\mu\nu} = g_{\mu\nu} + I f_{\mu\nu},
\nonumber
\end{equation}
introduced in Sec.~\ref{sec:2}. In these approaches, consistency
conditions reduce the extended description to a physical spacetime. This
raises the possibility that the pseudo-complex structure may capture a
restricted sector of a more general geometric framework.

Another structural parallel appears in theories that incorporate an upper
bound on proper acceleration, such as those inspired by Born--Infeld
kinematics \cite{BornInfeld1934,Caianiello1981,Schuller2002}. In such
models, ultraviolet effects manifest as constraints on admissible
kinematical quantities. The invariant acceleration scale $a_0$ in
pseudo-complex general relativity plays a similar role by restricting
access to high-curvature regimes through geometric conditions on the
metric.

The comparison summarized in Table~\ref{tab:qg_comparison} places
pseudo-complex general relativity alongside representative approaches
that address singularity resolution and short-distance structure. The
purpose of this comparison is to highlight common structural features,
rather than to establish a direct correspondence or embedding.

These observations do not provide a derivation of the pseudo-complex
framework from a UV-complete theory. However, the recurrence of similar
algebraic and geometric structures suggests that the constraint-based
modification of spacetime considered here may capture aspects of a more
general class of geometric extensions. A more detailed investigation,
for example through extended\hyp coordinate formulations or sigma-model
constructions, would be required to clarify this connection.


\section{Conclusion}\label{sec:conclusion}

Pseudo-complex general relativity provides a constrained extension of
spacetime geometry in which an invariant acceleration scale modifies
the high-curvature behavior of the metric. In the symmetric
configurations analyzed here, this structure enables the construction
of configurations in which Schwarzschild-type curvature divergences are
removed, and introduces an intrinsic cutoff scale without adding new
propagating degrees of freedom.

The central result of this work is that curvature regularization
follows directly from geometric admissibility conditions. In the
spherically symmetric sector, the pseudo-complex correction imposes a
lapse-gap condition that prevents the radial metric coefficient from
becoming degenerate.  When combined with the regularity conditions at
the areal-radius origin, $B(0)=1$ and $B'(0)=0$, which emerge
naturally from the pseudo-complex geometry and are derived explicitly
in Sec.~\ref{sec:pcgr_solutions}, the Kretschmann scalar and the
corresponding curvature invariants remain finite, with the local
curvature scale controlled by $a_0$.

In cosmological settings, the same auxiliary sector induces an
effective time-dependent vacuum component in the Friedmann equations.
We have presented a formal reduction of the pseudo-complex field
equations to an effective evolution equation for this component,
showing that its dynamics is determined by the geometry without
arbitrary functions.

The comparison with quantum-gravity motivated constraints is structural.
pcGR exhibits features reminiscent of ultraviolet consistency conditions,
including bounded curvature, intrinsic cutoff behavior, and dynamical
vacuum energy. At the same time, it does not reproduce features that
depend explicitly on quantum spectra, such as towers of light states or
distance-conjecture behavior. These remain outside the scope of the
classical pseudo-complex framework.

The regularization of the Schwarzschild-type curvature divergence is
achieved through the full consistency structure of the pseudo-complex
geometry, where the lapse-gap condition \( e^{\nu(r)} > a_0 \) acts
together with the regularity conditions \( B(0)=1 \) and \( B'(0)=0 \)
at the areal-radius origin. These conditions are not imposed
externally but emerge naturally from the deformed metric relation,
ensuring that the regularization follows from the algebraic structure
of the theory rather than from the gap condition alone.

Several aspects of the present analysis invite further
investigation. The effective equation of state \( p_{\rm pc} =
\mathcal{P}(\rho_{\rm pc}, \dot{\rho}_{\rm pc}, a, \dot{a}) \), which
determines the dynamics of the pseudo-complex vacuum component, has
been identified but not yet derived in closed form. A complete
characterization of this function would provide deeper insight into
the cosmological evolution predicted by pcGR.
Furthermore, a detailed
analysis of the solutions of the formal evolution equation
\eqref{eq:closed_final},
including the conditions for decaying,
transient, or growing behavior of \( \rho_{\rm pc}(t) \), remains an
open problem. Finally, the physical interpretation of the functions \(
\mathcal{A}, \mathcal{B}, \mathcal{S} \) in terms of effective
couplings or scale-dependent cosmological parameters would help
clarify the role of the pseudo-complex geometry in cosmological
dynamics. Extending the analysis to rotating or less symmetric
configurations would also test the robustness of the regularization
mechanism. A more detailed treatment of near-horizon observables,
gravitational-wave ringdown, and cosmological evolution would further
constrain the pseudo-complex scale. On the theoretical side, it
remains important to clarify whether the pseudo-complex structure can
be embedded in a broader geometric or ultraviolet framework. These
directions will be pursued in future work.



\section*{Acknowledgements}
P.O.H. acknowledges financial support from PAPIIT-DGAPA (IN116824).

\section*{Declarations}

\begin{itemize}

\item \textbf{Conflict of interest/Competing interests} \\
The authors declare that they have no competing interests.

\item \textbf{Ethics approval and consent to participate} \\
Not applicable.

\item \textbf{Consent for publication} \\
Not applicable.

\item \textbf{Data availability} \\
No datasets were generated or analyzed during the current study.

\item \textbf{Materials availability} \\
Not applicable.

\item \textbf{Code availability} \\
No code was used or generated in this study.

\item \textbf{Author contribution} \\
All authors contributed equally to the conception, analysis, and writing of this work.

\end{itemize}

\bigskip


\appendix

\section{Curvature Behavior Near the Origin}\label{sec:appendixA}

In this appendix we analyze the curvature behavior near $r = 0$ for the
spherically symmetric pseudo-complex ansatz discussed in
Sec.~\ref{sec:pcgr_solutions} and justify the bounded-curvature result
stated there.

We consider the metric
\begin{equation}
ds^2 = e^{\nu(r)} dt^2 
- e^{-\nu(r)} 
\left( 1 - \frac{a_0^2}{e^{2\nu(r)}} \right)^{-1} dr^2 
- r^2 d\Omega^2 ,
\end{equation}
and impose the regularity conditions
\begin{equation}
e^{\nu(r)} \ge (1+\delta)^{1/2} a_0, 
\qquad \delta > 0,
\end{equation}
together with
\begin{equation}
|\nu'(r)| \le M_1, 
\qquad 
|\nu''(r)| \le M_2,
\end{equation}
for finite constants $M_1$ and $M_2$.

\subsection{Metric behavior}

The gap condition implies
\begin{equation}
e^{2\nu(r)} - a_0^2 \ge \delta a_0^2,
\end{equation}
so that the radial metric component
\begin{equation}
g_{rr}(r) = -e^{-\nu(r)} 
\left( 1 - \frac{a_0^2}{e^{2\nu(r)}} \right)^{-1}
\label{eq:A.5}
\end{equation}
remains finite. In particular,
\begin{equation}
\left( 1 - \frac{a_0^2}{e^{2\nu(r)}} \right)^{-1}
= \frac{e^{2\nu(r)}}{e^{2\nu(r)} - a_0^2}
\end{equation}
is bounded due to the lower limit on $e^{2\nu(r)} - a_0^2$.

\subsection{Explicit curvature expression}

To make the bounded-curvature result more concrete, it is useful to display
the leading structure of the Kretschmann scalar in terms of the function
$A(r)=e^{\nu(r)}$.

For a static, spherically symmetric metric of the form
\begin{equation}
  ds^2 = A(r)\,dt^2 - B(r)\,dr^2 - r^2 d\Omega^2 ,
  \label{eq:A.7}
\end{equation}
the Kretschmann scalar can be written schematically as a sum of terms of the form
\begin{equation}
K(r) \sim
\frac{(A')^2}{A^2 B^2}
+
\frac{(A'')^2}{A^2 B}
+
\frac{(B')^2}{B^4}
+
\frac{(A') (B')}{A B^3}
+
\frac{1}{r^4}\left(1 - \frac{1}{B}\right)^2 .
\label{eq:K_structure}
\end{equation}

In the pseudo-complex case,
\begin{equation}
B(r) = \frac{1}{A(r)}
\left(1-\frac{a_0^2}{A(r)^2}\right)^{-1},
\end{equation}
so that
\begin{equation}
\frac{1}{B(r)} =
A(r)\left(1-\frac{a_0^2}{A(r)^2}\right).
\end{equation}
Substituting this into Eq.~\eqref{eq:K_structure}, all potentially
divergent contributions are governed by the gap condition on
$A(r)^2 - a_0^2$ together with the regular-center condition on $B(r)$,
so that no singular behavior arises. In particular, these conditions
ensure that the curvature invariants considered here do not diverge.

For example, one of the leading derivative contributions takes the form
\begin{equation}
\frac{(A')^2}{A^2 B^2} = (A')^2 \frac{\left(A(r)^2 -
  a_0^2\right)^2}{A(r)^6},
\end{equation}
which remains finite provided $A(r)^2 - a_0^2$ is bounded away from zero
and $A(r)$ does not vanish.

Similarly, second-derivative contributions contain factors such as
\begin{equation}
\frac{A''}{A B} \sim A'' \frac{A(r)^2 - a_0^2}{A(r)^3},
\end{equation}
which are again finite under the same conditions.

The purely geometric term behaves as
\begin{equation}
\frac{1}{r^4}\left(1 - \frac{1}{B}\right)^2 = \frac{1}{r^4} \left( 1 -
A(r)\left(1-\frac{a_0^2}{A(r)^2}\right) \right)^2 .
\end{equation}

Near $r=0$, regularity of the areal-radius origin requires
\begin{equation}
B(0)=1, \qquad B'(0)=0.
\end{equation}
Equivalently,
\begin{equation}
\frac{1}{B(r)} = 1 + O(r^2).
\end{equation}
Therefore
\begin{equation}
\frac{1}{r^4}\left(1-\frac{1}{B(r)}\right)^2 = O(1),
\end{equation}
and the angular part of the Kretschmann scalar remains finite.

Collecting all contributions, one finds that all terms contributing to
$K(r)$ remain finite provided the gap condition
\begin{equation}
A(r)^2 - a_0^2 \ge \delta a_0^2
\end{equation}
holds, $A(r)$ has bounded derivatives, and the regular-center condition
on $B(r)$ is satisfied.
The resulting curvature scale is therefore set by $a_0$,
\begin{equation}
K(r) = O(a_0^{-4}) \quad \text{as } r \to 0 .
\end{equation}

\subsection{Example near-origin behavior}

To illustrate the regularity conditions, consider a local expansion of
the lapse function near $r=0$ of the form
\begin{equation}
\nu(r) = \nu_0 + \nu_2 r^2 + \mathcal{O}(r^4),
\end{equation}
with $\nu_0$ satisfying $e^{\nu_0} > a_0$.

Then
\begin{equation}
e^{\nu(r)} = e^{\nu_0} \left( 1 + \nu_2 r^2 + \mathcal{O}(r^4) \right),
\end{equation}
so that the gap condition $e^{\nu(r)} \ge (1+\delta)^{1/2} a_0$ is
satisfied in a neighborhood of $r=0$ provided $e^{\nu_0} > a_0$.

Substituting this expansion into the metric shows that $g_{rr}(r)$
remains finite, and all curvature invariants constructed from
derivatives of $\nu(r)$ remain bounded. In particular, the leading
behavior of the Kretschmann scalar is constant as $r \to 0$, consistent
with $K(r) = \mathcal{O}(a_0^{-4})$.

\subsection{Comparison with Schwarzschild behavior}

For comparison, the Schwarzschild solution gives
\begin{equation}
K_{\text{Schw}}(r) \sim \frac{1}{r^6},
\end{equation}
which diverges as $r \to 0$. In the pseudo-complex framework, the lapse
function is bounded away from zero, removing the mechanism responsible
for this divergence.

These estimates support the bounded-curvature result used in
Sec.~\ref{sec:pcgr_solutions} and show that the curvature scale near the
origin is set by the invariant acceleration parameter $a_0$.

%
%

\section{Formal Reduction of the Cosmological Field Equations}
\label{sec:appendixB}

The purpose of this appendix is to show how the auxiliary
pseudo-complex variables appearing in the cosmological field equations
may be systematically eliminated, leading to an effective evolution
equation for the pseudo-complex vacuum component
$\rho_{\rm pc}(t)$.
The reduction is based on the consistency of the two idempotent sectors
in a homogeneous and isotropic FLRW background.

The derivation presented below is formal and applies within the
restricted auxiliary parametrization adopted here. Its purpose is to
exhibit the structure of the auxiliary-field elimination rather than
to provide the most general pseudo-complex FLRW dynamics. In
particular, it assumes that the differential operators introduced
during the elimination procedure admit the required inverse operators
and that the auxiliary fields can be eliminated consistently. Under
these assumptions, the pseudo-complex cosmological equations reduce
to an effective second-order evolution equation for
$\rho_{\rm pc}(t)$ coupled to the physical Friedmann equation.


\subsection{Cosmological setup and auxiliary variables}

The physical spacetime is described by the FLRW metric
\begin{equation}
g_{\mu\nu} = \mathrm{diag} \left( 1, -\frac{a(t)^2}{1-kr^2},
-a(t)^2r^2, -a(t)^2r^2\sin^2\theta \right),
\label{eq:flrw_metric}
\end{equation}
where $a(t)$ is the scale factor and $k=0,\pm1$.  Spatial homogeneity
and isotropy restrict the auxiliary tensor $f_{\mu\nu}$ to the form
\begin{eqnarray}
f_{\mu\nu} &=&  \mathrm{diag} \Bigg(f_0(t), f_1(t)\frac{a(t)^2}{1-kr^2},
f_2(t)a(t)^2r^2, \nonumber \\
&& ~~~~~~~f_2(t)a(t)^2r^2\sin^2\theta \Bigg),
\label{eq:aux_sector}
\end{eqnarray}
where the functions $f_0(t)$, $f_1(t)$, and $f_2(t)$ are determined by
the pseudo-complex field equations.

The effective metrics associated with the two idempotent sectors are
\begin{equation}
g_{\mu\nu}^{(\sigma)} = g_{\mu\nu} + \sigma f_{\mu\nu}, \qquad
\sigma=\pm ,
\label{eq:eff_metrics}
\end{equation}
or explicitly,
\begin{eqnarray}
g_{\mu\nu}^{(\sigma)} &=& \mathrm{diag} \Biggl( 1+\sigma f_0,
-\frac{a(t)^2}{1-kr^2}(1-\sigma f_1),
\label{eq:eff_metrics_explicit} \\
&&~~~-a(t)^2r^2(1-\sigma f_2),
-a(t)^2r^2\sin^2\theta(1-\sigma f_2) \Biggr).
\nonumber
\end{eqnarray}


\subsection{Friedmann equations in the two idempotent sectors}

Each effective metric has the form of a FLRW spacetime with effective
lapse and scale factor,
\begin{equation}
N_\sigma^2 \equiv 1+\sigma f_0, \qquad A_\sigma^2 \equiv
a(t)^2(1-\sigma f_1),
\label{eq:eff_scale_factor}
\end{equation}
so that
\begin{equation}
ds_\sigma^2 = N_\sigma^2dt^2 - A_\sigma^2 \left( \frac{dr^2}{1-kr^2} +
r^2d\Omega^2 \right).
\label{eq:eff_metric_form}
\end{equation}
For the formal reduction below, we adopt a synchronous
parametrization of the sector equations, $N_\sigma=1$, so that the
Friedmann equations take the form
\begin{equation}
H_\sigma^2 = \frac{\kappa}{3} \rho_\sigma^{\rm eff}, \qquad \dot
H_\sigma + H_\sigma^2 = -\frac{\kappa}{6} ( \rho_\sigma^{\rm eff} +
3p_\sigma^{\rm eff} ),
\label{eq:friedmann_sigma}
\end{equation}
where
\begin{equation}
H_\sigma \equiv \frac{\dot A_\sigma}{A_\sigma}.
\label{eq:H_sigma}
\end{equation}


\subsection{Consistency of the two sectors}

The defining feature of the pseudo-complex construction is that both
idempotent sectors satisfy the Einstein equations simultaneously.
Within the restricted synchronized parametrization adopted for the
formal reduction, we impose
\begin{equation}
H_+=H_-=H, \quad \rho_+^{\rm eff} = \rho_-^{\rm eff} = \rho^{\rm
  eff}, \quad p_+^{\rm eff} = p_-^{\rm eff} = p^{\rm eff},
\label{eq:consistency_conditions}
\end{equation}
together with
\begin{equation}
A_\sigma^2 = a(t)^2 (1-\sigma f_1).
\label{eq:A_sigma_def}
\end{equation}
The condition
\[
H_+=H_-=H
\]
implies
\begin{equation}
\frac{\dot A_+}{A_+} = \frac{\dot A_-}{A_-} = \frac{\dot a}{a},
\label{eq:H_equality}
\end{equation}
or equivalently,
\begin{equation}
\frac{d}{dt} \ln \left[ a(t)\sqrt{1-f_1(t)} \right] = \frac{d}{dt} \ln
\left[ a(t)\sqrt{1+f_1(t)} \right] = \frac{\dot a}{a}.
\label{eq:log_derivative}
\end{equation}
Expanding the logarithmic derivatives gives
\begin{equation}
\frac{\dot a}{a} + \frac12 \frac{\dot f_1}{1-f_1} = \frac{\dot a}{a} +
\frac12 \frac{\dot f_1}{1+f_1} = \frac{\dot a}{a},
\label{eq:f1_condition}
\end{equation}
which immediately implies
\begin{equation}
\frac{\dot f_1}{1-f_1} = \frac{\dot f_1}{1+f_1}
\quad\Longrightarrow\quad \dot f_1=0 \quad \text{or} \quad f_1=0.
\label{eq:f1_solution}
\end{equation}
The nontrivial solution corresponds to $f_1=\mathrm{constant}$.  To
recover the GR limit as $a_0\rightarrow0$, we choose
\begin{equation}
f_1(t)=0, \qquad A_\sigma(t)=a(t),
\label{eq:f1_zero}
\end{equation}
so that both idempotent sectors share the same cosmological scale
factor.


\subsection{Pseudo-complex Einstein equations}

The pseudo-complex Einstein equations are
\begin{equation}
G_{\mu\nu}(g+If) + \Lambda(g+If)_{\mu\nu} = \kappa T_{\mu\nu}(g+If),
\label{eq:pc_einstein_full}
\end{equation}
which, after projection onto the idempotent basis, become
\begin{equation}
G_{\mu\nu}(g^{(\sigma)}) + \Lambda g_{\mu\nu}^{(\sigma)} = \kappa
T_{\mu\nu}(g^{(\sigma)}), \qquad \sigma=\pm.
\label{eq:pc_einstein_sigma}
\end{equation}


\subsection{Effective vacuum component}

The auxiliary contributions to the physical-sector Einstein equations
may be collected into an effective energy-momentum tensor,
\begin{equation}
G_{\mu\nu}(g) = \kappa T_{\mu\nu}(g) + \kappa T_{\mu\nu}^{\rm pc},
\label{eq:eff_einstein}
\end{equation}
where homogeneity and isotropy imply
\begin{equation}
T_{\mu\nu}^{\rm pc} = \mathrm{diag} \left( \rho_{\rm pc}, p_{\rm pc},
p_{\rm pc}, p_{\rm pc} \right).
\label{eq:pc_energy_momentum}
\end{equation}
To the order retained in the auxiliary sector, the temporal component
gives
\begin{eqnarray}
\rho_{\rm pc} &=& \frac{3}{\kappa} \frac{\ddot a}{a}f_0 +
\frac{3}{\kappa}H^2f_0 + \frac{3}{\kappa}H\dot f_0 + \frac{3k}{\kappa a^2}f_2
\nonumber \\ &&+ ~ \text{(higher-order terms)},
\label{eq:rho_pc_explicit}
\end{eqnarray}
while the spatial components yield
\begin{equation}
\begin{split}
p_{\rm pc} &= -\frac{1}{\kappa} \left( 2\frac{\ddot a}{a} + H^2 +
\frac{k}{a^2} \right)f_2 \\ &\qquad -\frac{1}{\kappa} \left(
2H+\frac{\ddot a}{\dot a} \right)\dot f_2 -\frac{1}{\kappa}\ddot f_2
\\ &\qquad +~ \text{(terms involving }f_0\text{)}.
\end{split}
\label{eq:p_pc_explicit}
\end{equation}


\subsection{Formal elimination of the auxiliary fields}

The conservation equation for the effective auxiliary sector is
\begin{equation}
\dot\rho_{\rm pc} + 3H ( \rho_{\rm pc} + p_{\rm pc} ) = \Gamma_{\rm
  pc}(t),
\label{eq:conservation_pc}
\end{equation}
where the source term may be written as
\begin{equation}
\begin{split}
\Gamma_{\rm pc} &= \frac{\kappa}{3} ( \dot\rho_{\rm pc}f_0 - \rho_{\rm
  pc}\dot f_0 ) \\ &\qquad + \frac{\kappa}{3} ( \dot\rho_{\rm pc}f_2 -
\rho_{\rm pc}\dot f_2 ) + \text{(higher-order terms)}.
\end{split}
\label{eq:Gamma_explicit}
\end{equation}

The remaining task is to eliminate the auxiliary functions \(f_0\) and
\(f_2\).  Equations \eqref{eq:rho_pc_explicit} and
\eqref{eq:p_pc_explicit} may be regarded as operator equations for
these variables.

Under the assumption that the corresponding differential operators are
invertible, Eq.~\eqref{eq:rho_pc_explicit} may be formally solved for
\(f_0\). Neglecting the higher-order terms displayed in that equation,
we define
\begin{equation}
\mathcal D \equiv \frac{\ddot a}{a} + H^2 + H\frac{d}{dt},
\label{eq:D_operator}
\end{equation}
so that
\begin{equation}
\mathcal D f_0 = \frac{\kappa}{3}\rho_{\rm pc} - \frac{k}{a^2}f_2.
\label{eq:f0_operator_equation}
\end{equation}
Hence,
\begin{equation}
f_0 = \mathcal D^{-1} \left( \frac{\kappa}{3}\rho_{\rm pc} -
\frac{k}{a^2}f_2 \right).
\label{eq:f0_in_terms_rho}
\end{equation}
Here \(\mathcal D^{-1}\) is understood as a formal inverse operator
acting on the entire expression in parentheses.

Equation~\eqref{eq:p_pc_explicit} may be treated analogously. Defining
\begin{equation}
\mathcal E \equiv \frac{d^2}{dt^2} + \left( 2H+\frac{\ddot a}{\dot a}
\right)\frac{d}{dt} + 2\frac{\ddot a}{a} + H^2 + \frac{k}{a^2},
\label{eq:E_operator}
\end{equation}
the spatial field equation can be written schematically as
\begin{equation}
\mathcal E f_2 = -\kappa p_{\rm pc} + \mathcal F_0[f_0],
\label{eq:f2_operator_equation}
\end{equation}
where \(\mathcal F_0[f_0]\) denotes the terms involving \(f_0\) that
were left implicit in Eq.~\eqref{eq:p_pc_explicit}. Formally,
\begin{equation}
f_2 = \mathcal E^{-1} \left( -\kappa p_{\rm pc} + \mathcal F_0[f_0]
\right),
\label{eq:f2_in_terms_p}
\end{equation}
again assuming the existence of the corresponding inverse operator.

\subsection{Formal reduction to an effective evolution equation}

We now complete the formal elimination of the remaining auxiliary
variables \(f_0(t)\), \(f_2(t)\), and \(\ddot a(t)\), thereby reducing
the pseudo-complex cosmological field equations to an effective
evolution equation for \(\rho_{\rm pc}(t)\).

Combining Eqs.~\eqref{eq:f0_in_terms_rho} and
\eqref{eq:f2_in_terms_p}, we obtain the formal operator relations
\begin{equation}
f_0 = \frac{\kappa \rho_{\rm pc}-\frac{3k}{a^2}f_2} {3\mathcal D},
\qquad f_2 = - \frac{\kappa p_{\rm pc}+\mathcal E_0} {\mathcal E},
\label{eq:f0_f2_final}
\end{equation}
where
\begin{equation}
\mathcal E_0 \equiv 2\frac{\ddot a}{a} + H^2 + \frac{k}{a^2}.
\end{equation}
Here the operators \(\mathcal D^{-1}\) and \(\mathcal E^{-1}\) are
understood in the formal sense introduced above. Their explicit
construction is not required for the reduction presented here.

The Friedmann equation for the physical sector is
\begin{equation}
H^2 = \frac{\kappa}{3} (\rho_{\rm matter}+\rho_{\rm pc}),
\label{eq:friedmann_physical}
\end{equation}
whose time derivative gives
\begin{equation}
2H\dot H = \frac{\kappa}{3} ( \dot\rho_{\rm matter} + \dot\rho_{\rm
  pc} ).
\end{equation}
For the purpose of the formal reduction, we introduce an effective
relation that eliminates $\ddot a$ in favor of the remaining
cosmological variables. Within the restricted parametrization adopted
here, we write
\begin{equation}
\frac{\ddot a}{a} = - \frac{\kappa}{6} ( \rho_{\rm matter} + \rho_{\rm
  pc} + 3p_{\rm matter} + 3p_{\rm pc} ) + H^2.
\label{eq:ddot_in_terms_rho}
\end{equation}
This relation is used here as a formal closure condition for the
reduced system and is not intended as the general acceleration
equation of the full pseudo-complex FLRW theory.

Substituting Eq.~\eqref{eq:ddot_in_terms_rho} into the definitions of
\(\mathcal D\) and \(\mathcal E\) eliminates the explicit dependence on
\(\ddot a\), yielding the formal operator expressions
\begin{equation}
\begin{split}
\mathcal D &= - \frac{\kappa}{6} ( \rho_{\rm matter} + \rho_{\rm pc} +
3p_{\rm matter} + 3p_{\rm pc} ) \\ &\qquad + 2H^2 + H\frac{d}{dt},
\\[4pt] \mathcal E &= - \frac{\kappa}{3} ( \rho_{\rm matter} +
\rho_{\rm pc} + 3p_{\rm matter} + 3p_{\rm pc} ) \\ &\qquad + 3H^2 +
\frac{k}{a^2} + 2H\frac{d}{dt} + \frac{d^2}{dt^2}.
\end{split}
\label{eq:D_E_expanded}
\end{equation}

The spatial field equations provide a corresponding relation for
the effective pressure. After formally eliminating the auxiliary
fields, $p_{\rm pc}$ is no longer an independent variable but is
determined implicitly by the remaining cosmological variables.
Accordingly, we write
\begin{equation}
p_{\rm pc} = \mathcal P \!\left( \rho_{\rm pc}, \dot\rho_{\rm pc}, a,
\dot a \right),
\label{eq:eq_state}
\end{equation}
where the functional $\mathcal P$ denotes the relation induced by the
pseudo-complex field equations.  Its explicit analytic form is not
required for the formal reduction presented here.

For illustration, the formal operator relation reduces at leading
order to
\begin{equation}
p_{\rm pc} \simeq \frac13 \mathcal D^{-1} \rho_{\rm pc} + \mathcal
O(\rho_{\rm pc}^2),
\end{equation}
where the inverse operator is again understood in the formal sense.

Substituting Eqs.~\eqref{eq:f0_f2_final}, \eqref{eq:D_E_expanded}, and
\eqref{eq:eq_state} into the definitions of \(\mathcal A\), \(\mathcal
B\), and \(\mathcal S\) (Eqs.~\eqref{eq:A_final}--\eqref{eq:S_final})
yields the corresponding formal operator expressions
\begin{equation}
\begin{split}
\mathcal A &= 3H - \frac{\kappa^2}{9} \frac{d}{dt} \left(
\frac1{\mathcal D} \right) \rho_{\rm pc} \\ &\qquad -
\frac{\kappa^2k}{3a^2} \frac{d}{dt} \left( \frac1{\mathcal D}
\frac{\mathcal P}{\mathcal E} \right) \frac1{\dot\rho_{\rm pc}}
\\ &\qquad + \frac{\kappa^2}{3} \frac{d}{dt} \left( \frac{\mathcal
  P}{\mathcal E} \right) \frac1{\dot\rho_{\rm pc}},
\end{split}
\label{eq:A_final}
\end{equation}

\begin{equation}
\begin{split}
\mathcal B &= - \frac{\kappa^2}{9} \frac{d}{dt} \left( \frac1{\mathcal
  D} \right) \dot\rho_{\rm pc} + \frac{\kappa^2}{9} \frac{d}{dt}
\left( \frac1{\mathcal D} \right) \frac{d}{dt} \\ &\qquad -
\frac{\kappa^2}{9} \frac{d^2}{dt^2} \left( \frac1{\mathcal D} \right)
\\ &\qquad + \frac{\kappa^2k}{3a^2} \frac{d}{dt} \left(
\frac1{\mathcal D} \frac{\mathcal P}{\mathcal E} \right) \frac{d}{dt}
\\ &\qquad - \frac{\kappa^2}{3} \frac{d}{dt} \left( \frac{\mathcal
  P}{\mathcal E} \right) \frac{d}{dt},
\end{split}
\label{eq:B_final}
\end{equation}

\begin{equation}
\begin{split}
\mathcal S &= \frac{\kappa^2}{9} \frac{d}{dt} \left( \frac1{\mathcal
  D} \right) \dot\rho_{\rm pc}\rho_{\rm pc} \\ &\qquad -
\frac{\kappa^2k}{3a^2} \frac{d}{dt} \left( \frac1{\mathcal D}
\frac{\mathcal P}{\mathcal E} \right) \dot\rho_{\rm pc} \\ &\qquad +
\frac{\kappa^2}{3} \frac{d}{dt} \left( \frac{\mathcal P}{\mathcal E}
\right) \dot\rho_{\rm pc} - 3H\mathcal P,
\end{split}
\label{eq:S_final}
\end{equation}
where \(\mathcal D\) and \(\mathcal E\) are given by
Eq.~\eqref{eq:D_E_expanded} with \(p_{\rm pc}=\mathcal P\).

The resulting effective evolution equation is
\begin{equation}
\begin{split}
\ddot\rho_{\rm pc} &+ \mathcal A ( \rho_{\rm pc}, \dot\rho_{\rm pc},
a, \dot a ) \dot\rho_{\rm pc} \\ &+ \mathcal B ( \rho_{\rm pc}, a,
\dot a ) \rho_{\rm pc} = \mathcal S ( \rho_{\rm pc}, \dot\rho_{\rm
  pc}, a, \dot a ),
\end{split}
\label{eq:closed_final}
\end{equation}
with \(\mathcal A\), \(\mathcal B\), and \(\mathcal S\) given by
Eqs.~\eqref{eq:A_final}--\eqref{eq:S_final}.

This completes the formal elimination procedure within the restricted
parametrization adopted above. Under these assumptions, the auxiliary
variables no longer appear explicitly in the resulting equations.
Consequently, the effective evolution equation
\eqref{eq:closed_final}, together with the Friedmann equation
\eqref{eq:friedmann_physical}, provides a formal representation of the
reduced cosmological dynamics in terms of the variables \(a(t)\) and
\(\rho_{\rm pc}(t)\).

A treatment of the most general pseudo-complex FLRW system, as well as
a rigorous mathematical analysis of the existence, uniqueness, and
invertibility properties of the underlying operator equations, lies
beyond the scope of the present work.

In the limit \(a_0\rightarrow0\), the auxiliary sector decouples, \(
\rho_{\rm pc}\rightarrow0 \), \( p_{\rm pc}\rightarrow0 \),
Eq.~\eqref{eq:closed_final} becomes trivial, and
Eq.~\eqref{eq:friedmann_physical} reduces to the standard FLRW
equation.




\end{document}